\documentclass[12pt]{article}

\usepackage{amssymb,amsfonts,amsthm}
\usepackage{amsmath}
\usepackage{bm}             
\usepackage[mathscr]{eucal}
\usepackage{cancel}
\usepackage{wasysym}

\def\be{\begin{equation}}
\def\ee{\end{equation}}
\def\bea{\begin{eqnarray}}
\def\eea{\end{eqnarray}}

\def\bdelta{\mbox{\boldmath $\delta$}}

\def\hsp5{\hspace{5mm}}

\theoremstyle{remark}

\newcommand{\cA}{\mathcal{A}}

\newcommand{\cH}{\mathcal{H}}

\newcommand{\sfrac}[2]{{\textstyle{#1\over#2}}}

\title{\sc Asymptotic analysis of perturbed dust cosmologies to second order}

\begin{document}

\author{ \\
{\Large\sc Claes Uggla}\thanks{Electronic address:
{\tt claes.uggla@kau.se}} \\[1ex]
Department of Physics, \\
University of Karlstad, S-651 88 Karlstad, Sweden
\and \\
{\Large\sc John Wainwright}\thanks{Electronic address:
{\tt jwainwri@uwaterloo.ca}} \\[1ex]
Department of Applied Mathematics, \\
University of Waterloo,Waterloo, ON, N2L 3G1, Canada \\[2ex] }

\date{}
\maketitle

\begin{abstract}

Nonlinear perturbations of Friedmann-Lemaitre cosmologies with dust and a
cosmological constant $\Lambda>0$ have recently attracted considerable
attention. In this paper our first goal is to compare the evolution of the
first and second order perturbations by determining their asymptotic
behaviour at late times in ever-expanding models. We show that in the
presence of spatial curvature $K$ or a cosmological constant, the density
perturbation approaches a finite limit both to first and second order, but
the rate of approach depends on the model, being power law in the scale
factor if $\Lambda>0$ but logarithmic if $\Lambda=0$ and $K<0$. Scalar
perturbations in general contain a growing and a decaying mode. We find,
somewhat surprisingly, that if $\Lambda>0$ the decaying mode does not die
away, {\it i.e.}  it contributes on an equal footing as the growing mode to
the asymptotic expression for the density perturbation. On the other hand,
the future asymptotic regime of the Einstein-de Sitter universe ($K=\Lambda
=0$) is completely different, as exemplified by the density perturbation
which diverges; moreover, the second order perturbation diverges
\emph{faster} than the first order perturbation, which suggests that the
Einstein-de Sitter universe is unstable to perturbations, and that the
perturbation series do not converge towards the future. We conclude that the
presence of spatial curvature or a cosmological constant stabilizes the
perturbations. Our second goal is to derive an explicit expression for the
second order density perturbation that can be used to study the effects of
including a cosmological constant and spatial curvature.

\end{abstract}

\centerline{\bigskip\noindent PACS numbers: 04.20.-q, 98.80.-k, 98.80.Bp,
98.80.Jk}

\section{Introduction}

In a recent paper Uggla and Wainwright (2013) (referred to as
UW3)\footnote{The present paper also relies heavily on Uggla and Wainwright
(2011) and Uggla and Wainwright (2012), which we shall refer to as UW1 and
UW2, respectively.} we gave a simplified structure for the system of
equations that govern second order cosmological perturbations. Because of
their generality these equations provide a starting point for determining the
behaviour of nonlinear perturbations of Friedmann-Lemaitre (FL) cosmologies
with any given stress-energy content, using either the Poisson gauge or the
uniform curvature gauge. In the present paper we use this system of
equations, specialized to the Poisson gauge, to analyze the behaviour of
linear and nonlinear perturbations of FL cosmologies with dust and possibly a
cosmological constant as the matter-energy content. This problem has a
lengthy history starting with Tomita (1967), and has received considerable
attention over the past fifteen years.\footnote{See for example, Matarrese
{\it et al} (1998), Bartolo {et al} (2005), Tomita (2005), Boubekeur {\it et
al} (2008), Bartolo {\it et al} (2010) and  Hwang {\it et al} (2012).}

Our first goal is to compare the evolution in time of the nonlinear (second
order) perturbations to the linear (first order) perturbations by determining
the asymptotic behaviour of the perturbations at late times in ever-expanding
models, which has received little attention.\footnote{The only
papers of which we are aware are Bruni {\it et al} (2002) and Mena {\it et
al} (2002). They consider non-linear perturbations of a flat FL universe with
dust and a positive cosmological constant.  Their approach is heuristic in
that they solve a truncated asymptotic version of the evolution equation
for the perturbations in the synchronous gauge, whereas we solve the
general evolution equation in the Poisson gauge. In addition we focus
on the matter density perturbation at second order whereas they deal
with the metric perturbations.}
Since we do not wish to make the customary assumption that the spatial
geometry of the background is flat,\footnote{Current observations restrict
$\Omega_k$ to be close to zero. See, for example, Okouma {\it et al} (2013).
However, since a flat FL model is unstable to perturbations of $\Omega_k$
away from zero it is unlikely that $\Omega_k$ will remain close to zero. Thus
we are interested in how the presence of spatial curvature affects the future
asymptotic behaviour of perturbations.} there are three cases to consider:
\begin{itemize}
\item[i)] $\Lambda>0, K$ arbitrary, which we refer to as the de Sitter asymptotic regime,
\item[ii)] $\Lambda=0, K<0$, which we refer to as the Milne asymptotic regime,
\item[iii)] $\Lambda=0, K=0$, which we refer to as the Einstein-de Sitter asymptotic  regime.
\end{itemize}

We show that in the de Sitter and Milne asymptotic regimes, the second order
perturbations have the same asymptotic time dependence as the first order
perturbations. For example, in both cases the density perturbation "freezes
in", {\it i.e.} it approaches a finite limit as $x\rightarrow \infty$, where
$x=a/a_0$ is the dimensionless background scale factor:
\begin{equation} \lim_{x\rightarrow \infty}{}^{(r)}\!{\bdelta}=
{}^{(r)}\!{\bdelta}_{\infty},
\end{equation}
where $r=1,2$ labels the first order and second order perturbations,
respectively. There are two significant differences, however. First, in the
de Sitter regime, the rate of approach to the asymptotic state is power law
in $x$, while in the Milne regime it is logarithmic:
\be {}^{(r)}\!{\bdelta}_{de Sitter}=  {}^{(r)}\!{\bdelta}_{\infty}
+O(x^{-1}), \quad {}^{(r)}\!{\bdelta}_{Milne}=  {}^{(r)}\!{\bdelta}_{\infty}
+O(x^{-1} \ln x), \ee
as $x\rightarrow \infty$. Second, the limiting expression
${}^{(r)}\!{\bdelta}_{\infty}$ depends on the regime: in the de Sitter regime
it depends on both the growing and decaying modes of the linear perturbation,
while in the Milne regime it depends only on the growing mode at first order. In other words,
\emph{in the de Sitter asymptotic regime the decaying mode does not die
away}.

Finally, we show that the Einstein-de Sitter asymptotic regime is totally
different from the other two regimes. This is typified by the density
perturbation, which diverges:
\be ^{(1)}\!{\bdelta}_{EdS}=O(x), \quad ^{(2)}\!{\bdelta}_{EdS}=O(x^2), \quad
\text{as}\quad x\rightarrow\infty. \ee
In particular, the unbounded growth of these physical perturbations in the
Einstein-de Sitter asymptotic regime suggests that \emph{the Einstein-de
Sitter cosmology is unstable to perturbations}, and that the perturbation
series will not converge for $x$ sufficiently large. On the other hand our
analysis of the de Sitter and Milne asymptotic regimes as described above,
shows that \emph{the presence of spatial curvature or of a cosmological
constant stabilizes the perturbations.}

Our second goal is to derive an explicit expression for the second order
density perturbation that can be used to study the effects of including a
cosmological constant and spatial curvature. In order to obtain a relatively
simple expression which nevertheless illustrates various interesting
phenomena, we consider the special case in which the decreasing mode of the
linear perturbation is assumed to be zero, as is usually done in cosmological
perturbation theory. When the spatial curvature is set to zero our expression
is related to that given recently by Bartolo {\it et al} (2010).

The outline of the paper is as follows. In section 2 we give a new and
concise derivation of the solutions of the equations that govern first order
and second order perturbations, expressing them in integral form. Our
derivation is made simple by our use of the factorization property of the
linear differential operator that governs the evolution.\footnote{See UW1
equation (55) and UW2 equation (41).} We show that if the decaying mode of
the scalar perturbation is set to zero the general solution can be given in
explicit form, which leads to the expression for the density perturbation
referred to above. In section 3 we determine the asymptotic behaviour of the
perturbations in the three asymptotic regimes, and draw the conclusions
described above. Section 4 contains a brief summary and discussion. The
details of the analysis of the source terms and of the asymptotic behaviour
are given in the appendix.

\section{Perturbed dust cosmologies to second order}  \label{sec:simpleex}

In this section we solve the governing equations for linear and second order
perturbations of dust cosmologies, using the Poisson gauge. We make the
simplifying assumption that \emph{the perturbations at the linear order are
purely scalar}. This implies that the matter has zero vorticity.\footnote
{See equation (B.41d) in UW1.}
\subsection{Background}

We consider perturbations of a FL cosmology containing pressure-free matter
(dust) and a cosmological constant $\Lambda\geq 0$. The background
Robertson-Walker metric is given by
\be ds^2=a^2(-d\eta^2 +\gamma_{ij}dx^idx^j),  \ee
where $\eta$  is conformal time, $a$ is the scale factor and $\gamma_{ij}$ a
3-metric of constant curvature with curvature index $K$. The scale factor $a$
determines the dimensionless background Hubble scalar $\cal H$ according to
\be {\cal H}=\frac{a'}{a}=aH, \ee
where $H$ is the true background Hubble scalar and a prime denotes a
derivative with respect to $\eta$. As in UW3 we introduce the scalars
\be {\cal A}_G=2({\cal H}' -{\cal H}^2 + K), \quad {\cal A}_T=a^2
({}^{(0)}\!\rho + {}^{(0)}\!p), \ee
where
\be {}^{(0)}\!\rho = {}^{(0)}\!\rho_m + \Lambda  , \quad  {}^{(0)}\!p=
{}^{(0)}\!p_m  -\Lambda, \ee
where the subscript $m$ refers to the matter component. We assume that the
Einstein equations in the background are satisfied, which imply that ${\cal
A}_G={\cal A}_T$. We thus drop the subscripts $G$ and $T$, and for
pressure-free matter we have
\be \label{cal_A}   {\cal A} =2({\cal H}' -{\cal H}^2 + K)=a^2\,{}^{(0)}\!\rho_m. \ee
We introduce a reference time $\eta_0$ and a normalized scale factor $x$
defined by
\be \label{x_def}  x:=a/a_0, \quad \text{where} \quad a_0=a(\eta_0). \ee
For a dust source the matter
conservation equation in the background yields $(x^3{}^{(0)}\!\rho)'=0$,
which implies that  $({\mathcal A}x)'=0$. We define a constant $m$ by
\be \label{const_m} m^2:= \sfrac13{\mathcal A}x. \ee
We also introduce the usual density parameters:
\be \label{Omega} \Omega_m:=\frac{^{(0)}\!\rho_m}{3H^2}, \quad
\Omega_k:=-\frac{K} {{\cal H}^2}, \quad
\Omega_{\Lambda}:=\frac{\Lambda}{3H^2},  \ee
and note that for dust
\be  \label{cal_A2}  {\cal A}=3{\cal H}^2 \Omega_m, \quad m^2 = {\mathcal H}_0^2\,\Omega_{m,0}.  \ee
The Friedmann equation in the background, $\Omega_m +\Omega_k +\Omega_{\Lambda}=1$, can now be
used to express $\mathcal H$ explicitly as a function of $x$:
\be \label{H(x)}  {\mathcal H}^2 = {\mathcal H}_0^2\left(\Omega_{\Lambda,0}x^2+
\Omega_{k,0}+\Omega_{m,0}x^{-1}\right), \ee
with the subscript $0$ indicating evaluation at $\eta_0$.
%

\subsection{The governing equations}

In the Poisson gauge the gauge invariants that describe the metric
perturbation  are the \emph{Bardeen curvature} $^{(r)}\!\Psi$ and the
\emph{Bardeen potential} $^{(r)}\!\Phi$, and those that describe the matter
perturbation in the case of pressure-free matter (dust) and a cosmological
constant are $^{(r)}\!{\bdelta}$, the density perturbation, and $^{(r)}\!{\bf
v}$, the velocity perturbation, where $r=1$ at first order and $r=2$ at
second order. We refer to UW3 for the definition of $^{(r)}\!\Psi$ and
$^{(r)}\!\Phi$ (see equation (25)), and to equations~\eqref{gi}
and~\eqref{scalar_v} in the appendix of this paper for the definitions of
$^{(r)}\!{\bdelta}$ and $^{(r)}\!{\bf v}$. In UW3, in order to write the
governing equations for first and second order perturbations in a so-called
minimal form we also defined matter gauge invariants  in terms of the
components of the stress-energy tensor, denoted by ${}^{(r)}\!{\mathbb D}$
and ${}^{(r)}\!{\mathbb V}$, which we give in equations~\eqref{bbVD} in the
appendix.

We now specialize the governing equations derived in UW3~\footnote{See
equations (61).} to the case of dust. The Bardeen curvature $^{(r)}\!\Psi$
plays a central role in determining the dynamics of the perturbations. At
each order it satisfies a second order linear differential equation, which we
write in operator notation in the following form:
\begin{subequations} \label{cal_L}
\be \label{evol_psi} {\cal L}^{(1)}\!\Psi = 0, \qquad  {\cal L}^{(2)}\!\Psi
={\cal S}, \ee
where the differential operator ${\cal L}$ is given by
\be {\cal L}=\partial_{\eta}^2 + 3{\cal H}\partial_{\eta} + (2{\cal H}' +{\cal H}^2 -K), \ee
\end{subequations}
and the source term ${\cal S}$ depends quadratically on $^{(1)}\!\Psi$ and
its derivatives. We thus have to solve these equations successively, with the
solution of the first equation determining the source term on the right side
of the second equation. The remaining perturbation variables
$^{(r)}\!{\mathbb D},\, ^{(r)}\!{\mathbb V}$ and $ {}^{(r)}\!\Phi$, with
$r=1,2$,  are determined by $^{(1)}\!\Psi$ and $^{(2)}\!\Psi$ either
algebraically or by differentiation, as follows:
\begin{subequations} \label{constraint_DVPhi}
\begin{xalignat}{2}
^{(1)}\!{\mathbb D}& = 2{\cal A}^{-1}({\bf D}^2 + 3K){}^{(1)}\!\Psi,&\quad
^{(2)}\!{\mathbb D} &= 2{\cal A}^{-1}({\bf D}^2 + 3K){}^{(2)}\!\Psi  + {\cal S}_{\mathbb D}, \label{bbD2} \\
{\cal H}^{(1)}\!{\mathbb V} &= -2 {\cal A}^{-1}{\cal H}(\partial_{\eta} +
{\cal H}){}^{(1)}\!\Psi,&\quad {\cal H}^{(2)}\!{\mathbb V} &= -2 {\cal
A}^{-1}{\cal H}(\partial_{\eta} +
{\cal H}){}^{(2)}\!\Psi  + {\mathcal S}_{\mathbb V}, \label{bbV2} \\
{}^{(1)}\!\Phi &= {}^{(1)}\!\Psi, &\quad {}^{(2)}\!\Phi &= {}^{(2)}\!\Psi
+{\cal S}_{\Phi}, \label{phi2}
\end{xalignat}
\end{subequations}
where ${\bf D}^2 = \gamma^{ij}{\bf D}_i{\bf D}_j$ and ${\bf D}_i$ is the
spatial covariant derivative associated with $\gamma_{ij}$. The complete
expressions for the source terms ${\cal S}, {\cal S}_{\mathbb D},{\cal
S}_{\mathbb V}$ and ${\cal S}_{\Phi}$, which depend quadratically on
${}^{(1)}\!\Psi$ and its derivatives, are given in equations~\eqref{source}
in the appendix. Once ${}^{(r)}\!{\mathbb D}$ and ${}^{(r)}\!{\mathbb V}$
have been calculated one can obtain ${}^{(r)}\!{\bdelta}$ and ${}^{(r)}\!{\bf
v}$ by using the following relations, that are a special case of
equations~\eqref{matter_relations} in the appendix:
\begin{subequations} \label{rel_gen}
\begin{xalignat}{2}
{}^{(1)}\!{\bdelta} &= {}^{(1)}\!{\mathbb D} + 3{\cH}^{(1)}\!{\bf v}, &\quad
{}^{(2)}\!{\bdelta} &= {}^{(2)}\!{\mathbb D} + 3{\cH}^{(2)}\!{\mathbb V} -
2({\bf D}^{(1)}\!{\bf v}
)^2, \label{(2)delta} \\
{}^{(1)}\!{\bf v} &= {}^{(1)}\!{\mathbb V}, &\quad ^{(2)}\!{\bf v} &=
{}^{(2)}\!{\mathbb V} - 2{\cal S}^i\left[({}^{(1)}\!{\bdelta} -
{}^{(1)}\!\Psi){\bf D}_i( {}^{(1)}{\bf v})\right].
\end{xalignat}
\end{subequations}

For our purposes it is important that the operator ${\cal L}$ in
equations~\eqref{cal_L} can be written
as the product of two first order differential operators:
\begin{subequations} \label{op2}
\be \label{L_factor}  {\cal L}(\bullet) = {\cal H}{\cal L}_A{\cal L}_B\left(\frac{\bullet}{{\cal H}}\right), \ee
where
\be {\cal H}{\cal L}_A := {\cal H}\partial_\eta + 2{\cal H}' + {\cal H}^2, \qquad
{\cal L}_B := \partial_\eta + 2{\cal H}, \ee
\end{subequations}
(see UW1, equation (55) and UW2, equation (39)). Replacing conformal time
$\eta$ by the normalized scale factor $x$, noting that
$\partial_\eta={\mathcal H}x\partial_x$, we can now write ${\cal L}$ in the
form\footnote{Write the operators ${\mathcal L}_A, {\mathcal L}_B$ in the
form ${\cal H}{\mathcal L}_A(\bullet)=\partial_x({\cal H}^2 x\,\bullet)$ and
${\mathcal L}_B(\bullet)=\frac{{\cal H}}{x}\partial_x( x^2\,\bullet)$.}
\be \label{L_product} {\mathcal L}(\bullet)=\partial_x
\left({\mathcal H}^3\,\partial_x\!
\left(\frac{x^2}{\mathcal H}\bullet\right)\right). \ee
%

\subsection{Linear perturbations}

The first order Bardeen curvature $^{(1)}\!\Psi$ is obtained by solving the
first of equations~\eqref{evol_psi}, $ {\cal L}{}^{(1)}\!\Psi= 0$. We can
use~\eqref{L_product} to integrate  this equation twice, obtaining the
general solution in the form:\footnote{We will henceforth drop the
superscript $^{(1)}$ on first order quantities whenever there is no
danger of confusion.}
\begin{subequations}
\be\label{lin_Psi1} \Psi(x,x^i)= \frac{{\mathcal
H}}{x^2}\left(C_{+}(x^i)I(x)+C_{-}(x^i)\right), \ee
where
\be \label{I_x}  I(x) :=\int_0^x \frac{d{\bar x}}{{\mathcal H}({\bar x})^3}. \ee
\end{subequations}

The spatial functions $C_+(x^i)$ and $C_{-}(x^i)$ describe the growing and
decaying modes of the perturbation, respectively. The function $C_+(x^i)$ is
of particular significance, for the following reason. Two "conserved
quantities" that are associated with scalar perturbations of FL  have been
defined in the literature. These quantities, often denoted by the symbol
$\zeta$ with a subscript, satisfy an evolution equation of the form
\be \label{cons} \partial_\eta \zeta_{ \,\bullet} = {\bf D}^2
{\bf C}_{\,\bullet},  \ee
where ${\bf C}_{\,\bullet}$ is an expression involving the primary gauge
invariants such as $\Psi$ or ${\mathbb V}$ and the background variables. This
equation is referred to as a "conservation law", since if spatial derivatives
are negligible ("perturbations outside the horizon") in some epoch,
then~\eqref{cons} is approximated by $\partial_{\eta}\zeta_{ \,\bullet} = 0$,
{\it i.e.} $\zeta_{ \,\bullet}$ is approximately constant in time during that
epoch. We refer to UW2, section 4, for a unified discussion of these
conserved quantities. In particular the conserved quantity that we denote by
$\zeta_{\mathrm v}$ can be expressed in terms of $\Psi$ and ${\mathbb V}$
according to\footnote{See UW2, equation (72). This quantity is sometimes
referred to as "the curvature perturbation in the comoving gauge". See Malik
and Wands (2009), equation (7.46), who use the symbol ${\cal R}$ for
$\zeta_{\mathrm v}$.}
\be \label{zeta}  \zeta_{\mathrm v} = \left( 1 -\frac{2K}{\cA}
\right)\Psi - {\cH}{\mathbb V}. \ee
On substituting for ${\mathbb V}$ from the first of equations~\eqref{bbV2}
we can rearrange~\eqref{zeta} to obtain
\be  \zeta_{\mathrm v} = \frac{2{\cal H}^3}{{\cal A}x}\partial_x \left(\frac{x^2}{\cal H}\Psi\right). \ee
Substituting for $\Psi$ from~\eqref{lin_Psi1} and using~\eqref{const_m} leads to
\be \zeta_{\mathrm v} = \sfrac23 m^{-2}C_{+}. \ee
This result shows firstly that for any perturbation with dust, {\it the
conserved quantity $\zeta_{\mathrm v}$ is exactly constant in time}, and
secondly it relates the arbitrary spatial function $C_{+}(x^i)$ to
$\zeta_{\mathrm v}$. We can thus write the general solution~\eqref{lin_Psi1}
in the form
\be \label{(1)Psi_gen} \Psi(x,x^i)= \sfrac32 m^2\frac{{\mathcal H}}{x^2}
\left(I(x)\zeta(x^i)+C_{-}(x^i)\right), \ee
on rescaling $C_{-}(x^i)$. Here and in the rest of the paper, for notational
convenience we drop the subscript $\mathrm v$ on $\zeta_{\mathrm v}$. We note
in passing that in other discussions of scalar perturbations of dust, another
conserved quantity, the so-called curvature perturbation on constant density
hypersurfaces,\footnote{This quantity is denoted by $\zeta$ in Malik and
Wands (2009) (see equation (7.61)), and by $\zeta_{\rho}$ in UW2  (see
equations (65)-(67)). In the long wavelength regime $\zeta_{\rho}$ and
$\zeta_{\mathrm v}$ are approximately equal, but unlike $\zeta_{\mathrm v}$,
$\zeta_{\rho}$ is not exactly constant at first order.} plays an important
role (see, for example, Bartolo {\it et al} (2006), equation (2.13)).

The other gauge invariants can be expressed in terms of $\Psi$. First, the
matter gauge invariant $\mathbb{D}$ is given by the first of
equations~\eqref{bbD2}. Second, the gauge invariant $\mathbb{V}$ is given by
the first of equations~\eqref{bbV2}, or can be calculated using~\eqref{zeta}
since $\zeta$ is constant in time. Finally we have $\Phi=\Psi$ as follows
from the first of equations~\eqref{phi2}.

\subsection{Second order perturbations}

The second order Bardeen curvature $^{(2)}\!\Psi$ is obtained by solving the
second of equations~\eqref{evol_psi}, ${\cal L}{}^{(2)}\!\Psi= {\mathcal
S}(\Psi)$. The solution of this equation that satisfies the initial
conditions
\be  {}^{(2)}\!\Psi(x_{init},x^i)=0, \qquad \partial_x
{}^{(2)}\!\Psi(x_{init},x^i)=0,  \ee
where $x_{init}>0$, can be obtained by using \eqref{L_product} and integrating twice. This leads
to the following formula:
\begin{subequations}
\be \label{(2)Psi_gen}  {}^{(2)}\!\Psi(x,x^i)=\frac{{\mathcal H} }{x^2}
\int_{x_{init}}^x \frac{ {\mathbb S}({\bar x},x^i)}{ {\mathcal H}({\bar x})^3
}
 d{\bar x}, \ee
where we have defined
\be \label{bb_S_gen} {\mathbb S}(x,x^i):= \int_{x_{init}}^{ x} {\mathcal S}({\tilde x},x^i) d{\tilde x}. \ee
Here
\be \label{S_gen} {\mathcal S}(x, x^i) \equiv {\mathcal S}(\Psi(x,x^i)),  \ee
\end{subequations}
with $\Psi(x,x^i)$ given by~\eqref{(1)Psi_gen}.
The general solution can be written in the form
\be  \label{2Psi_gen}  {}^{(2)}\!\Psi_{gen} =  {}^{(2)}\!\Psi +
{}^{(2)}\!\Psi_{homog}, \ee
where ${}^{(2)}\!\Psi_{homog}$ is the general solution of the homogeneous
equation ${\cal L}{}^{(2)}\!\Psi=0$ and hence is of the
form~\eqref{lin_Psi1}, with $C_+(x^i)$ and $C_-(x^i)$ being determined by the
values of ${}^{(2)}\!\Psi_{gen}$ and $\partial_\eta{}^{(2)}\!\Psi_{gen}$ at
the initial time $x_{init}$.

The other perturbed quantities at second order, $^{(2)}\!{\mathbb D},
{}^{(2)}\!{\mathbb V}$ and $^{(2)}\!\Phi$, are determined explicitly in terms
of ${}^{(2)}\!\Psi$ and $^{(1)}\!\Psi$ by the constraint
equations~\eqref{constraint_DVPhi}. We now give an alternative expression for
$^{(2)}\!{\mathbb V}$. By differentiating~\eqref{(2)Psi_gen} with respect to
$\eta$, using $\partial_{\eta}={\mathcal H}x\partial_x$, one can express
$(\partial_{\eta} + {\cal H}){}^{(2)}\!\Psi$ in terms of ${\mathbb S}({
x},x^i)$ and ${}^{(2)}\!\Psi$. This leads to the following expression
\be \label{calc_V}   {\cal H}\, ^{(2)}\!{\mathbb V} = -\sfrac23
m^{-2}\,{\mathbb S}(x,x^i) + \left( 1 - \frac{2K}{\cal A}
\right){}^{(2)}\!\Psi +{\mathcal S}_{\Bbb V}  , \ee
which is useful for analyzing the asymptotic behaviour of ${}^{(2)}\!{\mathbb
V}$.

Finally we note that the solution~\eqref{(2)Psi_gen} can be written in an
alternate form by changing the order of integration, yielding
\be \label{(2)Psi_alt}   {}^{(2)}\!\Psi(x,x^i) = \frac{{\mathcal H} }{x^2}
\int_{x_{init}}^x [I(x)-I(\bar x)]{\mathcal S}(\bar x,x^i)d{\bar x}.  \ee
%

\subsection{Zero decaying mode at linear order}

The solution $^{(2)}\!\Psi$, given by~\eqref{(2)Psi_gen}
or~\eqref{(2)Psi_alt}, depends on the two arbitrary spatial functions $\zeta$
and $C_{-}$ through the source term ${\cal S}$ which depends on
$^{(1)}\!\Psi$. We can obtain detailed information about the time and spatial
dependence of ${}^{(2)}\!\Psi$ by considering the special case of a linear
perturbation with {\it zero decaying mode},\footnote{This simplifying
assumption is often made, usually because it is claimed that the decaying
mode will become negligible compared to the growing mode in the future.
However, as we show in Section~\ref{sec:lin_desitter}, this is not the case
if $\Lambda>0$: the decaying mode contributes to the perturbations at both
linear and second order. If one assumes that the decaying mode is zero one is
essentially considering perturbations in a universe with an isotropic
singularity (Goode and Wainwright (1985)).}
{\it i.e.} $C_{-}(x^i)=0$.

When the decaying mode is zero, equation~\eqref{(1)Psi_gen} and
the first of equations~\eqref{bbV2} reduce to
\be {\Psi}(x,x^i)= g(x)\zeta(x^i), \qquad  {\cal H}{\bf v}=-\sfrac23
\Omega_m^{-1} fg\,\zeta,\ee
\be {\bdelta} =\sfrac23g\left[x\left(\frac{{\bf D}^2}{m^2}\right) -3\Omega_m^{-1}(f+\Omega_k)  \right]\zeta,  \ee
where\footnote{We note in passing that the function $D_{+}:=xg(x)$, defined
up to a constant factor, is sometimes referred to as the growth suppression
factor. See for example Bartolo {\it et al} (2006), in the text following
equation (2.3). In our analysis the related function $I(x)$ plays a central
role.}
\be  \label{fg} f(x):=1 +\frac{xg'}{g}, \qquad g(x):= \sfrac32 m^2
\frac{{\mathcal H}I}{x^2}. \ee
We use~\eqref{(2)Psi_alt} to obtain a particular solution for
${}^{(2)}\!\Psi$. In this case one can if desired choose $x_{init}=0$
in~\eqref{(2)Psi_alt}, since ${\cal S}=O(1)$ as  $x\rightarrow\infty,$ as
follows from~\eqref{S_spec} and~\eqref{T_A}. We now substitute the
expression~\eqref{S_spec} for the source term ${\cal S}$
into~\eqref{(2)Psi_alt}, which leads to
\begin{subequations} \label{(2)Psi_special}
\be  {}^{(2)}\!\Psi(x,x^i)=\frac{1}{x}\left( { B}_1(x)\zeta
^2+ { B}_2(x) {\cal D}(\zeta)+ m^{-2}[{ B}_3(x)({\bf D}\zeta)^2
+ { B}_4(x){\bf D}^2 {\cal D}(\zeta)] \right), \ee
where
\be  \label{B_A}   {B}_A(x):= \frac{\cal H}{x}\int_{x_{init}}^x [I(x)-I(\bar x)]T_A(\bar x)d{\bar x}, \quad
A=1,\dots,4,  \ee
\end{subequations}
and the functions $T_A(x)$ are given by~\eqref{T_A}. The notation $({\bf
D}f)^2$ and the operator ${\cal D}$ are defined in~\eqref{DA}.

The density perturbation ${}^{(2)}\!{\bdelta}$ is given by~\eqref{(2)delta}.
We calculate the expression ${}^{(2)}\!{\mathbb D} +3{\cal H}{}^{(2)}\!{\mathbb V}$
that is required by substituting ${}^{(2)}\!{\Psi}$, as given
by~\eqref{(2)Psi_special}, into the second of equations~\eqref{bbD2}
and~\eqref{bbV2}. After using~\eqref{const_m},~\eqref{cal_A2} and
$\partial_{\eta}={\cal H}x\partial_x$ we obtain
\be {}^{(2)}\!{\mathbb D} +3{\cal H}{}^{(2)}\!{\mathbb V} =
-2\Omega_m^{-1}{\hat \partial}_x(x{}^{(2)}\!{\Psi}) +
\sfrac23 m^{-2}{\bf D}^2(x{}^{(2)}\!{\Psi}) +{\cal S}_{\mathbb D}+3 {\cal S}_{\mathbb V}. \ee
 The source terms ${\cal S}_{\mathbb D}$ and
${\cal S}_{\mathbb V}$ are given by~\eqref{source}. The final result is
\be \label{(2)delta_special} \begin{split} {}^{(2)}\!{\bdelta} = A_1 \zeta^2
+ A_2{\cal D}(\zeta) &+
m^{-2}\left[A_3({\bf D}\zeta)^2+A_4{\bf D}^2{\cal D}(\zeta)+A_5{\bf D}^2 \zeta^2 \right]  \\
& +\sfrac23m^{-4}\left[B_3{\bf D}^2({\bf D}\zeta)^2  + B_4{\bf D}^4{\cal D}(\zeta)  \right],
\end{split}\ee
where
\begin{subequations} \label{delta_coeff}
\begin{align}
A_1&=-2\Omega_m^{-1}\left[ {\hat \partial}_x B_1 - g^2\left((1-f)^2-4\Omega_k\right)\right],\\
A_2&=-2\Omega_m^{-1}\left[{\hat \partial}_x B_2 - 4g^2(1+\sfrac{2}{3}\Omega_m^{-1} f^2)\right], \\
A_3&=-2\Omega_m^{-1}\left[{\hat \partial}_x B_3 +\sfrac13 xg^2\left(5\Omega_m+\sfrac{4}{3} f^2\right)\right],\\
A_4&= -2\Omega_m^{-1}\left[{\hat \partial}_x B_4 -\sfrac13 \Omega_m{B}_2\right],  \quad
A_5=\sfrac23({B}_1 +4xg^2),
\end{align}
with
\be {\hat \partial}_x B_A:= ({\partial}_x + \Omega_k x^{-1})B_A.  \ee
\end{subequations}
Here the time-dependent functions $A_1$ and $A_2$ identify the
\emph{Newtonian terms} that dominate at late times, $A_3-A_5$ identify the
\emph{post-Newtonian terms}, while $B_3$ and $B_4$ identify the
\emph{super-horizon terms} that describe the perturbations on the largest
scales. These time-dependent functions depend on $\Lambda$ and $K$ through
the functions ${\cal H}(x)$ and $I(x)$, and when $K=0=\Lambda$ and we choose
$x_{init}=0$ they reduce to powers of $x$, as given in
section~\ref{sec:Eds1}. Observe that the spatial dependence is described by
expressions such as $({\bf D}\zeta)^2$ that are quadratic in $\zeta$ and its
spatial derivatives. In the super-horizon, post-Newtonian and Newtonian terms
these expressions are of degree zero, two and four in the spatial derivative
operator ${\bf D}$, respectively.

Equation~\eqref{(2)Psi_special} gives a particular solution for
${}^{(2)}\!\Psi$ that satisfies  $\lim_{x\rightarrow 0}{}^{(2)}\!\Psi=0$,
and~\eqref{(2)delta_special} gives the corresponding expression for the
density perturbation ${}^{(2)}\!{\bdelta}$. The general solution for
${}^{(2)}\!\Psi$ subject to the condition that the decaying mode at linear
order is zero is given by\footnote{The term ${}^{(2)}\!\Psi_{homog}$
in~\eqref{2Psi_gen} is of the form~\eqref{lin_Psi1} with $C_{-}=0$.}
\be \label{Psi_and_C}  {}^{(2)}\!\Psi_{gen} = {}^{(2)}\!\Psi + C(x^i)g(x),
\ee
where $^{(2)}\!\Psi$ is given by~\eqref{(2)Psi_special}.
The corresponding density perturbation is given by
\be  \label{bdelta_gen}  {}^{(2)}\!{\bdelta}_{gen}= {}^{(2)}\!\bdelta + \sfrac23 g\left[
x\,\frac{{\bf D}^2}{m^2}C -3\Omega_m^{-1}(f+\Omega_k) C\right],  \ee
where ${}^{(2)}\!\bdelta$ is given by~\eqref{(2)delta_special}.

The expression for $ {}^{(2)}\!{\bdelta}_{gen}$ given
by~\eqref{(2)delta_special},~\eqref{delta_coeff} and~\eqref{bdelta_gen} is
new and represents one of the main results of this paper. We have shown that
the expression for ${}^{(2)}\!\bdelta$  given by Bartolo {\it et al} (2010)
when $K=0$ and $\Lambda>0$ can be written in our form\footnote{See their
equation (29). Note that their $g(\eta)$ equals our $g(x)$ up to a constant
multiple: $g(x)=\sfrac35 g(\eta)$, and our $f(x)$ equals their $f(\eta)$.
Some rearrangement of the spatial dependence terms has to be done to relate
their functions $B_A(x^i)$ to ours.} provided that the spatial function
$C(x^i)$ is chosen suitably. Specifically, in deriving their result they use
the level of primordial non-Gaussianity at the end of inflation ({\it ibid},
section 3) to determine the function $C(x^i)$ in~\eqref{Psi_and_C}, which has
the form
\be \label{C_special}   C(x^i)= \sfrac43 g_{in}[2{\cal D(\zeta)}+(1 -\sfrac52a_{\mathrm nl})\zeta^2], \ee
where $a_{\mathrm nl}$ is a constant that parametrizes the primordial
non-Gaussianity and $g_{in}$ is the value of $g$ at some initial time.
Finally we note that Tomita (2005) has also given an expression for
${}^{(2)}\!{\bdelta}$ when  $K=0$ and $\Lambda>0$, in which the time
dependence is expressed in a completely different way.\footnote{See Tomita's
equation (4.16).}

\subsubsection{A special case: Einstein-de Sitter}  \label{sec:Eds1}

When $K=0$ and $\Lambda=0$, it follows from \eqref{H(x)} and~\eqref{I_x} that
${\cal H}$ and $I(x)$ are given by
\be  \label{asymp_m}  {\cal H}^2 =  m^2 x^{-1}, \qquad m^3
I(x)=\sfrac25\,x^{5/2}, \ee
and hence that the time dependence functions $f$ and $g$ are
\be g(x)= \sfrac35, \qquad f(x)=1. \ee
One can now use~\eqref{T_A} to calculate the functions $T_A(x)$, which when substituted
in~\eqref{B_A} with $x_{init}=0$ leads to
\be B_1=0=B_2, \qquad  B_3 =\sfrac{2}{175}x^2, \qquad B_4=20B_3. \ee
Equations~\eqref{(2)Psi_special},~\eqref{(2)delta_special} and~\eqref{delta_coeff} now lead to
\be \label{(2)Psi_m}   ^{(2)}\!\Psi ={\cal F}(\zeta)x,  \quad \text{with} \quad
{\cal F}(\zeta):=\sfrac{2}{175}m^{-2}[({\bf D}\zeta)^2 + 20 {\bf D}^2
{\cal D}(\zeta)],  \ee
\be  \label{(2)del_m}   {}^{(2)}\!{\bdelta} =\sfrac{24}{5}{\cal D}(\zeta)+
\sfrac{2}{175}m^{-2}\left(137({\bf D}\zeta)^2 - 80{\bf D}^2 {\cal D}(\zeta)
+84{\bf D}^2\zeta^2\right)x +\sfrac23m^{-2}{\bf D}^2{\cal F}(\zeta) x^2.  \ee
The corresponding general solution is given by~\eqref{Psi_and_C}
and~\eqref{bdelta_gen}. The second order density perturbation in the Poisson
gauge for the Einstein-de Sitter universe has been given by a number of
authors, including Matarrese {\it et al} (1998), (equation (6.10)), Bartolo
{\it et al} (2005) (equation (8)), Hwang {\it et al} (2012), (equation (43)).
We have shown that the expressions for ${}^{(2)}\!\bdelta$ given by these
authors can be written in our form with the arbitrary function $C(x^i)$ given
by~\eqref{C_special} with $g_{in}=\sfrac35$ and $a_{\mathrm nl}=0$ in the
first and third papers.

\section{Asymptotic behaviour at late times}

We now turn to the problem of determining the asymptotic behaviour of the
second order perturbations as $x\rightarrow \infty$, that arise from a
general  (scalar) linear perturbation~\eqref{(2)Psi_gen}, {\it i.e.} one that
has \emph{both} a growing mode $\zeta$ and a decaying mode $C_{-}$. It is
thus necessary to use the general solution~\eqref{(2)Psi_gen} for
${}^{(2)}\!\Psi$. We consider the three cases listed in the introduction.

\subsection{The de Sitter asymptotic regime}

In this section we determine the  form of the nonlinear perturbation in the
 de Sitter asymptotic regime: $x\rightarrow \infty$ with $\Lambda>0, K$
arbitrary (assuming background solutions that are forever expanding when
$K>0$). The first step is to use the asymptotic form of $^{(1)}\!\Psi$ to
determine the asymptotic form of the source term $ {\mathcal S}(\Psi)$
in~\eqref{S_gen}. Then the asymptotic behaviour of ${}^{(2)}\!\Psi$ can be
determined using equation~\eqref{(2)Psi_gen}.

\subsubsection{Linear perturbations} \label{sec:lin_desitter}

The asymptotic form of $\Psi$ is determined by the functions ${\cal H}(x)$
and $I(x)$ through equation~\eqref{(1)Psi_gen}. It follows from~\eqref{H(x)}
and~\eqref{I_x} that as $x\rightarrow \infty$
\begin{subequations} \label{asymp_l}
\begin{align} {\mathcal H} &= \lambda x\left(1+\sfrac12 k_{\lambda} x^{-2}
+O(x^{-3})\right), \quad \text{where}\quad \lambda := \sqrt{
\Omega_{\Lambda,0}{\mathcal H}_0^2}, \\
{\lambda}^3 I(x) &= I_{\infty}-\sfrac{1}{2}x^{-2} + O(x^{-4}), \quad \text{where} \quad
I_{\infty}:={\lambda^3}\int_0^{\infty} \frac{1}{{\cal H}(x)^3}dx. \label{I_infty}
\end{align}
\end{subequations}
Equation \eqref{(1)Psi_gen} then leads to
\begin{subequations}
\be  \label{(1)Psi_l} \Psi(x,x^i) = \sfrac32 m_{\lambda}
\,^{(1)}\!G_{\Lambda}(x^i)x^{-1} +O(x^{-3}),  \quad \text{as} \quad
x\rightarrow \infty, \ee
where
\be \label{(1)F_l}  ^{(1)}\!G_{\Lambda}(x^i):= I_{\infty}\zeta(x^i) + C_{-}(x^i). \ee
\end{subequations}
Here we have introduced the scaling parameters
\be \label{nu}   m_{\lambda}:=\frac{m^2}{\lambda^2}
=\frac{\Omega_{m,0}}{\Omega_{\Lambda,0}}, \qquad
k_{\lambda}:=-\frac{K}{\lambda^2}=\frac{\Omega_{k,0}}{\Omega_{\Lambda,0}}.
\ee
It follows from \eqref{zeta},~\eqref{bbD2}~\eqref{(1)Psi_l} and~\eqref{rel_gen} that
the asymptotic behaviour as $x\rightarrow\infty$ of the matter gauge invariants is given by
\begin{subequations}
\begin{align}
{}^{(1)}\!{\bdelta}&=m_{\lambda}\frac{{\bf D}^2}{m^2}
{}^{(1)}\!G_{\Lambda} - 3\zeta + O(x^{-1}),   \label{(1)D_l}  \\
{\cal H} {}^{(1)}\!{\bf v} &= -(\,\zeta -
k_{\lambda}^{(1)}\!G_{\Lambda}) + O(x^{-1}).   \label{(1)V_l}
\end{align}
\end{subequations}
%

\subsubsection{Second order perturbations}

A particular solution for ${}^{(2)}\!\Psi$ is given by~\eqref{(2)Psi_gen}. In
order to determine the asymptotic behaviour of  ${}^{(2)}\!\Psi$ as
$x\rightarrow\infty$ we need to determine the asymptotic behaviour of the
source term ${\cal S}(x, x^i)$. The key result is that ${\cal S}(x,
x^i)=f(x^i)+ O(x^{-2}).$ We now derive this result, obtaining an explicit
expression for the leading order spatial term $f(x^i)$.

It follows from~\eqref{S_simple} and the asymptotic forms of $^{(1)}\!\Psi$
and ${\cal H}^{(1)}\!{\mathbb V}$ (details in the appendix) that
\be \label{S_asymp_l}    {\mathcal S} = {\mathcal H}^2 \left[\Psi^2 +
 4{\cal D}(\Psi)\right] + O(x^{-2}).  \ee
Since ${\cal S}(x,x^i)$ is quadratic in $^{(1)}\!\Psi$ and its derivatives it
will have a multiplicative factor of $m_{\lambda}^2$. We thus rescale
${S}(x,x^i)$ and ${\mathbb S}(x,x^i)$ by defining
\be \label{scale_S_l}    {\mathcal S}(x,x^i) =
\sfrac32{\lambda}^2{m_{\lambda}^2}\,{\bar {\mathcal S}}(x,x^i), \qquad
{\mathbb S}(x,x^i) = \sfrac32{\lambda}^2{m_{\lambda}^2}\,{\bar {\mathbb
S}}(x,x^i). \ee
It now  follows from~\eqref{(1)Psi_l},~\eqref{S_asymp_l} and~\eqref{scale_S_l} that
\be \label{S_l}   {\bar {\mathcal S}}(x,x^i)={\cal G}_{\Lambda}(x^i) + O(x^{-2}), \ee
where
\be \label{calG_l}  {\cal G}_{\Lambda}(x^i):=\sfrac32
\left(^{(1)}\!G_{\Lambda}^2 + 4{\cal D}(^{(1)}\!G_{\Lambda})\right). \ee
This in turn implies that
\be \label{bbS_l}  {\bar {\mathbb S}}(x,x^i)={\cal G}_{\Lambda}(x^i)\,x + O(1). \ee

We now use~\eqref{(2)Psi_gen} and~\eqref{bbS_l} to obtain the following
asymptotic expansion for $ {}^{(2)}\!\Psi$ as $x\rightarrow\infty$ (details
in the appendix):
\begin{subequations}
\be \label{(2)Psi_l}   {}^{(2)}\!\Psi(x,x^i)=  \sfrac32 m_{\lambda}^2
\,\left({}^{(2)}\!G_{\Lambda}(x^i)x^{-1} - {\cal G}_{\Lambda}(x^i)x^{-2}
+O(x^{-3})\right), \ee
where
\be  \label{(2)G_l}   {}^{(2)}\!G_{\Lambda}(x^i) := \int_{x_{init}}^{\infty}
\frac{ {\bar{\mathbb S}}({\bar x},x^i)}{ \left({\sfrac{1}{\lambda}\mathcal
H}({\bar x})\right)^3 }d{\bar x}. \ee
\end{subequations}
This improper integral converges since ${\cal H}(x)=O(x)$ and ${\bar{\mathbb
S}}(x,x^i)=O(x)$ as $x\rightarrow\infty$.

The asymptotic behaviour as  $x\rightarrow\infty$ of the remaining second
order perturbation variables is determined by
equations~\eqref{constraint_DVPhi},~\eqref{source} and~\eqref{rel_gen}:
\begin{subequations}
\begin{align}
\label{asym_Phi_l} {}^{(2)}\!\Phi &= {}^{(2)}\!\Psi + O(x^{-2}), \\
\label{(2)del_l}  {}^{(2)}\!{\bdelta} &= m_{\lambda}^2 \frac{{\bf
D}^2}{m^2} {}^{(2)}\!G_{\Lambda} - 3m_{\lambda}{}^{(2)}\!\zeta +
O(x^{-1}), \\
\label{asym_V_l}   {\cal H}{}^{(2)}\!{\bf v} &= -m_{\lambda}\,\left(^{(2)}\!\zeta -
k_{\lambda}^{(2)}\!G_{\Lambda}\right) + {\cal S}_{\bf v} +O(x^{-1}),
\end{align}
where
\be {\cal S}_{\bf v}= 2{\cal S}^i\left[  \left(m_{\lambda} \frac{{\bf D}^2}{m^2} {}^{(1)}\!G_{\Lambda} -
3\zeta \right){\bf D}_i \left(\zeta -k_{\lambda}{}^{(1)}\!G_{\Lambda}\right) \right].   \ee
\end{subequations}
Here $^{(2)}\!\zeta$, as defined by~\eqref{(2)zeta}, is a quantity that
depends quadratically on the first order function $^{(1)}\!G_{\Lambda}(x^i)$
given by~\eqref{(1)F_l}. If  the decaying mode is zero ($C_{-}(x^i)=0$) then
$^{(2)}\!\zeta$ depends quadratically on $\zeta$, as does ${\cal S}_{\bf v}$.
In this special case one can also express $^{(2)}\!G_{\Lambda}$ and
${}^{(2)}\!\zeta $ as a linear combination of the terms quadratic in $\zeta$
and its spatial derivatives that appear in~\eqref{(2)Psi_special}
and~\eqref{(2)delta_special}.

\subsection{The Milne asymptotic regime }

In this section we determine the  form of the nonlinear perturbation in the
asymptotic to Milne regime: $x\rightarrow \infty$ with $\Lambda=0, K<0$.

\subsubsection{Linear perturbations}

It follows from \eqref{H(x)} and~\eqref{I_x} that the asymptotic  behaviour
of ${\cal H}$ and $I(x)$ as $x\rightarrow\infty$ is given by
\begin{subequations} \label{asymp_k}
\be {\mathcal H}= \sqrt{-K}\left(1+\sfrac12m_k x^{-1} +O(x^{-2})\right),
\quad \text{where} \quad m_k:=\frac{\Omega_{m,0}}{\Omega_{k,0}} =-\frac{m^2}{K},   \ee
\be (-K)^{\sfrac32} \,I(x)=x\,(1-\sfrac32m_k \frac{{\ln x}}{x} + O(x^{-1}) ). \ee
\end{subequations}
Equation \eqref{(1)Psi_gen} then leads to
\be  \label{(1)Psi_k} \Psi(x,x^i) = \sfrac32 m_k\,
\zeta(x^i)x^{-1}(1-\sfrac32 m_k \frac{\ln x}{ x}) +O(x^{-2}), \quad
\text{as} \quad x\rightarrow \infty. \ee
It follows from \eqref{zeta},~\eqref{bbD2},~\eqref{(1)Psi_k} and~\eqref{rel_gen} that
the asymptotic behaviour as $x\rightarrow\infty$ of the matter gauge invariants is given by\begin{subequations}
\begin{align}
{}^{(1)}\!{\bdelta} &= (m_k\, \frac{{\bf D}^2}{m^2}  -3)\zeta
+ O( x^{-1}\ln x), \label{(1)D_k}   \\
{\cal H} {}^{(1)}\!{\bf v} &= -\sfrac32 m_k\,\zeta\,
\frac{\ln x}{x} + O(x^{-1}).  \label{(1)V_k}
\end{align}
\end{subequations}
%

\subsubsection{Second order perturbations}

A particular solution for ${}^{(2)}\!\Psi$ is given by~\eqref{(2)Psi_gen}. In
order to determine the asymptotic behaviour of  ${}^{(2)}\!\Psi$ as
$x\rightarrow\infty$ we need to determine the asymptotic behaviour of the
source term ${\cal S}(x, x^i)$. The key result is that ${\cal S}(x,
x^i)=f(x^i)x^{-2}+ O(x^{-3} \ln x).$ We now derive this result, obtaining an
explicit expression for the leading order spatial term $f(x^i)$.

It follows from~\eqref{S_simple} and the asymptotic forms of $\Psi$ and
${\cal H}{\mathbb V}$ (details in the appendix) that
\be \label{S_asymp_k}  {\mathcal S} = K[3\Psi^2 +4{\cal D}(\Psi)] +\sfrac13[-({\bf D}\Psi)^2 +
4{\bf D}^2 {\mathcal D}(\Psi)] + O(x^{-3}(\ln x)^2).  \ee
Since ${ S}(x,x^i)$ is quadratic in $^{(1)}\!\Psi$ and its derivatives it
will have a multiplicative factor of $m_k^2$. We thus rescale ${S}(x,x^i)$
and ${\mathbb S}(x,x^i)$ by defining
\be \label{scale_S_k}   {S}(x,x^i) = \sfrac32(-K){m_k^2}\,{\bar {S}}(x,x^i),
\qquad {\mathbb S}(x,x^i) = \sfrac32(-K){m_k^2}\,{\bar {\mathbb S}}(x,x^i).
\ee
It now  follows from~\eqref{(1)Psi_k},~\eqref{S_asymp_k} and~\eqref{scale_S_k} that
\be \label{S_k}  {\bar S}(x,x^i)={\cal G}_{k}(x^i)x^{-2} + O(x^{-3}(\ln x)^2), \ee
where
\be \label{calG_k}  {\mathcal G}_{k}(x^i):= -\sfrac32\left[3\,\zeta^2 +
4{\cal D}(\zeta)  + \sfrac13 {m_k}m^{-2} \left(({\bf D}\zeta)^2 - 4{\bf
D}^2{\mathcal D}(\zeta)\right)\right].  \ee
This in turn implies that
\be\label{intS_k} \int_{x_{init}}^{\infty} {\bar{\cal S}}(\bar x,x^i) d{\bar
x} =\lim_{x\rightarrow \infty}{\bar{\mathbb S} }(x,x^i), \quad \text{exists
and is non-zero}.   \ee

We now use the general solution~\eqref{(2)Psi_gen} in conjunction
with~\eqref{intS_k} to obtain the following asymptotic expansion for
${}^{(2)}\!\Psi $ as $x\rightarrow\infty$ (details in the appendix):
\begin{subequations}
\be\label{(2)Psi_k} {}^{(2)}\!\Psi(x,x^i) = \sfrac32 m_k^2\,x^{-1}\left[
^{(2)}\!G_{k} - \left(\sfrac32 m_k \,^{(2)}\!G_{k} + {\cal G}_k
\right)\frac{\ln x}{ x}\right]+O(x^{-2}),  \ee
where
\be \label{(2)G_k} {}^{(2)}\!G_{k}(x^i):=  \int_{x_{init}}^{\infty}{\bar
{\cal S}}({\tilde x},x^i)d{\tilde x} . \ee
\end{subequations}

In the limit $x\rightarrow\infty$ the remaining second order perturbation
variables are determined in terms of ${}^{(2)}\!\Psi$ and ${}^{(1)}\!\Psi$ by
equations~\eqref{constraint_DVPhi},~\eqref{source},~\eqref{calc_V} and~\eqref{rel_gen}:
\begin{subequations}
\begin{align}  \label{asym_Phi_k} {}^{(2)}\!\Phi &= {}^{(2)}\!\Psi + O(x^{-2}), \\
\label{asym_D_k}    ^{(2)}\!{\bdelta} &= m_k(m_k \,\frac{{\bf D}^2}{m^2} -3){}^{(2)}\!G_k
 + O( x^{-1}\ln x),  \\
\label{asym_V_k}  {\cal H} ^{(2)}\!{\bf v} &= -\sfrac32 m_k^2\, {}^{(2)}\!G_k
\frac{\ln x}{x} + O(x^{-1}).
\end{align}
\end{subequations}
We note that if the decaying mode is zero one can express $^{(2)}\!G_k$ as a
linear combination of the terms quadratic in $\zeta$ and its spatial
derivatives that appear in~\eqref{(2)Psi_special}
and~\eqref{(2)delta_special}.

\section{Discussion}

We have used the minimal system of governing equations for second order
perturbations that we developed in UW3 to investigate the behaviour of
nonlinear scalar perturbations of FL cosmologies with dust and cosmological
constant as matter-energy content. Although we have given a new formulation
of the solutions of the governing equations, our main contribution in this
paper has been to analyze the properties of the solutions, and this has been
facilitated by the new form of the solutions. In particular we have shown
that there are significant differences in the asymptotic behaviour at late
times in three cases, characterized by $K$ and $\Lambda$, namely, the
Einstein-de Sitter regime, the Milne regime and the de Sitter regime, as
described in the introduction. These differences are best illustrated by
considering the behaviour of the linear and nonlinear density perturbations
${}^{(1)}\!{\bdelta}$ and ${}^{(2)}\!{\bdelta}$. For ease of comparison we
repeat the three asymptotic expressions for ${}^{(1)}\!{\bdelta}$ as
$x\rightarrow\infty$:
\begin{subequations} \label{delta_infty}
\begin{align}
^{(1)}\!{\bdelta}_{deSitter}&=m_{\lambda}\frac{{\bf D}^2}{m^2}
(I_{\infty}\zeta +C_{-}) - 3\zeta + O(x^{-1}),  \\
{}^{(1)}\!{\bdelta_{Milne}} &= m_k\, \frac{{\bf D}^2}{m^2}\zeta  -3\zeta
 + O( x^{-1}\ln x), \\
{}^{(1)}\!{\bdelta}_{EdS} &= \sfrac25\ x\frac{{\bf D}^2}{m^2}\zeta
 + O(1),
\end{align}
\end{subequations}
where $I_{\infty}$ is given by~\eqref{I_infty}. Note that the expressions for
the density perturbation $^{(2)}\!{\bdelta}$ in the three asymptotic regimes
are given by equations~\eqref{(2)del_l},~\eqref{asym_D_k}
and~\eqref{(2)del_m}.

An arbitrary scalar perturbation depends on two spatial functions $\zeta(x^i)$ and
$C_{-}(x^i)$, that correspond to the growing and decaying modes, and on the
state of the background model at some initial time, which determines the
following constants:
\be m^2={\cal H}_0^2 \,\Omega_{m,0}, \qquad
m_{\lambda}=\frac{\Omega_{m,0}}{\Omega_{\Lambda,0}}, \qquad k_{\lambda}=\frac{\Omega_{k,0}
}{\Omega_{\Lambda,0} }, \qquad m_k=\frac{\Omega_{m,0} }{ \Omega_{k,0}}.
\ee
It follows from~\eqref{I_infty} that
\be I_{\infty}=I_{\infty}(m_{\lambda},k_{\lambda}). \ee
By inspection of equations~\eqref{delta_infty} we see the following
dependencies of the asymptotic states for the first order density
perturbation ${}^{(1)}\!\bdelta$:
\begin{subequations}
\begin{align}
\text{Einstein-de Sitter}: &\quad m,\zeta, \\
\text{Milne}: &\quad m,m_k,\zeta,  \\
\text{de Sitter}: &\quad m,m_{\lambda},k_{\lambda},\zeta,C_{-}.
\end{align}
\end{subequations}
We see that ${}^{(1)}\!\bdelta$ has the most complicated structure in the de
Sitter asymptotic regime, depending on the ratio of $\Omega_{m,0}$ and
$\Omega_{k,0}$ to $ \Omega_{\Lambda,0}$ and on the decaying mode $C_{-}$. The
situation is different as regards the second order density perturbation.
Since ${}^{(2)}\!\Psi$ and hence ${}^{(2)}\!\bdelta$ has an integral
dependence on the source term ${\cal S}$ (see equation~\eqref{(2)Psi_alt}),
the decaying mode enters into the leading term in all three asymptotic
regimes.

\subsection*{Acknowledgments}
CU thanks the Department of Applied Mathematics at the University of Waterloo
for kind hospitality. JW acknowledges financial support from the University
of Waterloo.

\begin{appendix}

\section{Analysis for second order perturbations}

\subsection{Source terms for second order perturbations}

In this section we give the full expressions for the source terms and then
illustrate their structure in the special case when the decaying mode is
zero.

The source terms for a perfect fluid model with a purely scalar linear
perturbation are given in UW3 (see equations (61)). After specializing to the
case of dust and rearranging the terms we obtain\footnote{In UW3 the source
terms are not given a label ${\cal S}_{\bullet}$, but can be identified by
comparing equations~\eqref{evol_psi} and~\eqref{constraint_DVPhi} with
equations (61) in UW3, noting the difference in choice of variables as
described in equation~\eqref{old_new}. We have also used the first
equation in~\eqref{bbV2} in obtaining these expressions.}
\begin{subequations}  \label{source}
\begin{align}
\label{S_simple} \begin{split}
{\cal S} &=  \,({\cal H}x\,\partial_x\Psi)^2 + 4K\Psi^2  -
\sfrac13 \left(({\bf D}\Psi)^2 - {\cal A}({\bf D}{\mathbb V})^2 \right) \\
& \quad + 2\left[{\cal H}^2( x \partial_x +3\Omega_{\Lambda} + \Omega_k) +
\sfrac13{\bf D}^2  \right] \left(2{\cal D}(\Psi) + {\cal A}{\cal D}({\mathbb
V}) \right),
\end{split} \\
{\cal S}_{\mathbb D} &= 2{\cal A}^{-1}[4({\bf D}^2+3K)\Psi^2 -5({\bf D}\Psi)^2 ] +
6 {\cal S}^i(\Psi {\bf D}_i{{\cal H}\mathbb V}) + \sfrac32{\cal A} {\mathbb V}^2, \\
{\cal S}_{\mathbb V} &= 2{\cal A}^{-1}{\cal H}^2\left[\Psi^2 + 4{\cal D}(\Psi)\right] +
 2\left[{\cal S}^i({\cal H}{\mathbb V}{\bf D}_i\Psi) +2{\cal D}({\cal H}{\mathbb V}) \right], \label{S_V} \\
{\cal S}_{\Phi} &= 4[\Psi^2 - {\cal D}(\Psi)]  - 2{\cal A}{\cal D}({\mathbb V}).
\end{align}
\end{subequations}
Here and elsewhere we use the following notation. First, the second order
spatial differential operators are defined by
\begin{equation} \label{2order_D}
{\bf D}^2 := \gamma^{ij}{\bf D}_i{\bf D}_j, \qquad {\bf
D}_{ij} := {\bf D}_{(i}{\bf D}_{j)} - \sfrac13 \gamma_{ij}{\bf D}^2,
\end{equation}
where ${\bf D}_i$ denotes covariant differentiation
with respect to the spatial metric $\gamma_{ij}$.
Second,  we use the shorthand notation
\be \label{DA}  \left({\bf D}A\right)^2:=({\bf D}^k A)( {\bf D}_kA), \qquad
{\cal D}(A):={\mathcal S}^{ij} ({\bf D}_ i A)({\bf D}_j A), \ee
where $A$ is a scalar field. Finally,  we use the scalar mode extraction operators
\be\label{extract} {\cal S}^{i} = {\bf D}^{-2}{\bf D}^i ,\qquad {\cal S}^{ij}
= \sfrac32 {\bf D}^{-2}\!\left({\bf D}^2 + 3K\right)^{-1}{\bf D}^{ij},  \ee
as defined in UW3 (see equation (85a)).

\subsubsection{Zero decaying mode at linear order}

It follows from \eqref{S_simple} that ${\mathcal S}(x,x^i)$ can be written in the form
\be \label{S_spec}   {\mathcal S}(x,x^i)= T_1(x)\zeta^2+
T_2(x){\cal D}(\zeta) + m^{-2}\left(T_3(x)({\bf D}\zeta)^2 +
T_4(x){\bf D}^2 {\cal D}(\zeta)\right),   \ee
where
\begin{subequations} \label{T_A}
\begin{align}
T_1 &= ({\mathcal H}g)^2((f-1)^2- 4\Omega_k), \\\
T_2 &=  -8({\mathcal H}g)^2\left( (f-1)^2 - \sfrac12(3\Omega_{\Lambda} + \Omega_k) +
\sfrac{2\Omega_k}{3\Omega_m}f^2\right), \\
T_3 &= -\sfrac13m^2 g^2\left(1-\sfrac{4}{3\Omega_m}f^2 \right), \\
T_4 &= \sfrac43 m^2g^2\left(1+ \sfrac{2}{3\Omega_m}f^2\right),
\end{align}
\end{subequations}
and $f$ and $g$ are defined in~\eqref{fg}. The remaining source terms are
given by
\begin{subequations}  \label{S_spec_other}
\begin{align}
{\cal S}_{\mathbb D}&=\sfrac23 g^2\left[x m^{-2}\left(4{\bf D}^2\zeta^2 -5({\bf D}\zeta)^2\right)+
\sfrac{3}{\Omega_m}\left(f(f-1)-4\Omega_k\right)\zeta^2\right],  \\
{\cal S}_{\mathbb V}&=\sfrac{2}{3}\Omega_m^{-1}g^2\left[-f \zeta^2 + 4\left(1+\sfrac{2}{3\Omega_m}f^2\right) {\cal D}(\zeta)\right],  \\
{\cal S}_{\Phi}&=4g^2\left[\zeta^2 -\left(1+\sfrac{2}{3\Omega_m}f^2\right) {\cal D}(\zeta)\right].
\end{align}
\end{subequations}

\subsection{Asymptotic expressions for the source terms}
\subsubsection{Asymptotic to de Sitter}

In order to derive~\eqref{S_asymp_l} we have to examine~\eqref{S_simple} in
detail. We begin by using ~\eqref{H(x)},~\eqref{(1)Psi_l} and~\eqref{(1)V_l}
to conclude that ${\cal D}(\Psi)$ and ${\cal A}{\cal D}({\mathbb V})$ have
the asymptotic form
\be {\cal D}(\Psi)= A(x^i)x^{-2} + O(x^{-4})  ,\qquad  {\cal A}{\cal
D}({\mathbb V}) =  B(x^i)x^{-3} + O(x^{-4}),  \ee
and
\be \label{step_1}  {\cal H}^2( x \partial_x + 3\Omega_{\Lambda} + \Omega_k)
+ \sfrac13{\bf D}^2 = \lambda^2 x^2( x \partial_x +3) + O(1). \ee
It then follows that
\be \label{step_2}   x^2(x \partial_x +3) {\cal D}(\Psi)= x^2{\cal D}(\Psi)
+O(x^{-2}), \qquad  x^2(x \partial_x +3){\cal A}{\cal D}({\mathbb
V})=O(x^{-2}). \ee
The desired result~\eqref{S_asymp_l} follows when~\eqref{step_1}
and~\eqref{step_2} are substituted in~\eqref{S_simple}.\footnote{The terms in
the first line of~\eqref{S_simple} are ${\cal H}^2\Psi^2 + O(x^{-2})$.} The
asymptotic forms as $x\rightarrow\infty$ of the other source terms given
by~\eqref{source} are:
\begin{subequations} \label{source_l}
\begin{align}
{\mathcal S}_{\Phi} &= 9m_{\lambda}^2\left[G_{\Lambda}^2 - {\cal D}(G_{\Lambda})\right]x^{-2} +O(x^{-3}),  \\
{\mathcal S}_{\mathbb D} &=  O(x^{-1}), \\
\label{S_V_l1}  {\mathcal S}_{\mathbb V} &= m_{\lambda}{\mathcal G}_{\Lambda} x  + 4{\cal D}\left(\zeta -
k_{\lambda} G_{\Lambda} \right) +O(x^{-1}) .
\end{align}
\end{subequations}
The leading order term in ${\mathcal S}_{\mathbb D}$ is quite complicated,
and is not needed for our purposes.

\subsubsection{Asymptotic to Milne}

In order to derive~\eqref{S_asymp_k} we have to examine~\eqref{S_simple} in
detail. We begin by using ~\eqref{H(x)},~\eqref{(1)Psi_k}
and~\eqref{(1)V_k} to conclude that that ${\cal D}(\Psi)$ and ${\cal A}{\cal
D}({\mathbb V})$ have the asymptotic form
\be {\cal D}(\Psi)= f(x^i)x^{-2} + O(x^{-4}),\qquad {\cal A}{\cal D}({\mathbb
V})=  g(x^i)x^{-3}(\ln x)^2 + O(x^{-4}),  \ee
and
\be \label{step_1k}  {\cal H}^2( x \partial_x  + \Omega_k) + \sfrac13{\bf D}^2 =
(-K)( x \partial_x +1) + \sfrac13 {\bf D}^2 + O(x^{-2}). \ee
It then follows that
\be \label{step_2k}   (x \partial_x +1) {\cal D}(\Psi) =  -{\cal D}(\Psi) +
O(x^{-4}),   \qquad (x \partial_x +1){\cal A}{\cal D}({\mathbb V}) =
O(x^{-3}(\ln x)^2).  \ee
The desired result~\eqref{S_asymp_k} follows when~\eqref{step_1k}
and~\eqref{step_2k} are substituted in~\eqref{S_simple}.\footnote {The terms
in the first line of~\eqref{S_simple} are $3K\Psi^2 -\sfrac13({\bf D}\Psi)^2
+ O(x^{-3}\ln x)$.} The asymptotic forms as $x\rightarrow\infty$ of the other
source terms given by~\eqref{source}, are:
\begin{subequations} \label{source_k}
\begin{align}
{\mathcal S}_{\Phi} &= O(x^{-2}),  \\
{\mathcal S}_{\mathbb D} &= \sfrac32{m_k}^2m^{-2}\left[ 4({\bf D}^2 +
3K)\zeta^2 - 5({\bf D}\zeta)^2 \right]x^{-1} + O(x^{-2}{\ln x}), \\
\label{S_V_k} {\mathcal S}_{\mathbb V} &= \sfrac32m_k\,[\zeta^2 + 4{\cal
D}(\zeta)]x^{-1} + O(x^{-2}({\ln x})^2).
\end{align}
\end{subequations}
\subsubsection{Einstein-de Sitter}

The exact expressions for the source terms given by~\eqref{source} are
\begin{subequations}  \label{source_m}
\begin{align}
\label{S_m} {\mathcal S} &= \sfrac19[({\bf D}\Psi)^2+20{\bf D}^2 {\cal D}(\Psi)], \\
{\mathcal S}_{\Phi} &= 4[\Psi^2 -\sfrac53{\cal D}(\Psi)],  \\
{\mathcal S}_{\mathbb D} &= 2{\cal A}^{-1}[4{\bf D}^2 \Psi^2 -5({\bf D}\Psi)^2], \\
{\mathcal S}_{\mathbb V} &= \sfrac{40}{9}{\cal D}(\Psi).
\end{align}
\end{subequations}
\subsection{Asymptotic expansions for second order perturbations}
\subsubsection{Asymptotic to de Sitter}

We outline the derivation of equation~\eqref{(2)Psi_l} for $^{(2)}\!\Psi$. It
follows from~\eqref{(2)Psi_gen},~\eqref{nu} and~\eqref{scale_S_l} that
\be   x {}^{(2)}\!\Psi(x,x^i)= \sfrac32 m_{\lambda}^2 \frac{\cH}{\lambda x}
\int_{x_{init}}^{x} \frac{{\bar {\mathbb S}}(\bar x,x^i) }
{(\sfrac{1}{\lambda}{\mathcal H}({\bar x}))^3 }d{\bar x}. \ee
Since
\be \label{int1} {\bar{\mathbb S}}(x,x^i)=x{\cal G}_{\Lambda}(x^i) +O(1), \quad  \frac{\cH}{\lambda x}=1+ O(x^{-2}), \ee
(see equations~\eqref{H(x)} and~\eqref{bbS_l}) it follows that
$\lim_{x\rightarrow \infty} x {}^{(2)}\!\Psi(x,x^i)$ exists. With
$^{(2)}\!G_{\Lambda}(x^i)$ defined as in~\eqref{(2)G_l} we obtain
\be  x {}^{(2)}\!\Psi(x,x^i) -\sfrac32 m_{\lambda}^2{}^{(2)}\!G_{\Lambda}(x^i) = -
\int_{x}^{\infty} \frac{{\bar {\mathbb S}}(\bar x,x^i) }
{(\sfrac{1}{\lambda}{\mathcal H}({\bar x}))^3 }d{\bar x}. \ee
The integral on the right can be expanded using~\eqref{int1}, leading to the
desired equation~\eqref{(2)Psi_l}.

To calculate $^{(2)}\!{\mathbb V}$ we need to extend the asymptotic expansion
of ${\bar {\mathbb S}}$ given by~\eqref{bbS_l}. We write
\be
\label{bbS_l_extra}  {\bar {\mathbb S}}(x,x^i) = {\cal G}_{\Lambda}(x^i)\,x + {\bar {\mathbb S}}_0 + O(x^{-1}).
\ee
When~\eqref{bbS_l_extra} and~\eqref{S_V_l1} are substituted
into~\eqref{calc_V}, after making use of~\eqref{scale_S_l}, we obtain the
desired equation~\eqref{asym_V_l}, with
\be  \label{(2)zeta}   ^{(2)}\!\zeta:={\bar {\mathbb S}}_0 +
4{\cal D}\left(\zeta -
k_{\lambda} G_{\Lambda} \right). \ee
\subsubsection{Asymptotic to Milne}

We outline the derivation of equation~\eqref{(2)Psi_k} for $^{(2)}\!\Psi$. We
begin by using~\eqref{scale_S_l} to write~\eqref{(2)Psi_gen} in the form
\be  \label{(2)Psi_k1}    x {}^{(2)}\!\Psi(x,x^i)= \sfrac32 m_k^2
\frac{{\bar \cH}}{ x} \int_{x_{init}}^{x} \frac{{\bar {\mathbb S}}(\tilde
x,x^i) } {{\bar {\mathcal H}}({\tilde x})^3 }d{\tilde x}, \ee
where
\be \label{asymp1} {\bar \cH}(x) := (-K)^{-1/2}{\cH}(x)= (1+m_k
x^{-1})^{\sfrac12}. \ee
We next derive an asymptotic expansion for ${\bar {\mathbb S}}(x,x^i)$. It
follows from equations~\eqref{bb_S_gen}, \eqref{scale_S_k} and~\eqref{(2)G_k}
that
\be {\bar {\mathbb S}}(x,x^i) - ^{(2)}\!\!G_k(x^i) = - \int_{x}^{\infty}{\bar
{\cal S}}({\tilde x},x^i)d{\tilde x}.  \ee
Substituting the expansion ~\eqref{S_k} of ${\bar {\cal S}}({\tilde x},x^i)$
and evaluating the integral gives
\be \label{bbS_k}   {\bar {\mathbb S}}(x,x^i) = {}^{(2)}\!G_k(x^i) - {\cal
G}_k(x^i) x^{-1} + O(x^{-2} \ln x), \ee
which in conjunction with~\eqref{asymp1} leads to
\be  \frac{{\bar {\mathbb S}}(x,x^i) } {{\bar {\mathcal H}}( x)^3 } =
{}^{(2)}\!G_k - \left(\sfrac32 m_k \,{}^{(2)}\!G_k + {\cal G}_k\right)
x^{-1}+ O(x^{-2} \ln x).  \ee
On substituting this expression in~\eqref{(2)Psi_k1} and evaluating the
integral we obtain~\eqref{(2)Psi_k}.

\section{The matter gauge invariants}

In this appendix we define the two types of matter gauge invariants that we
use in this paper. We consider a perfect fluid with stress-energy tensor
\be \label{pf}  T^a\!_b = \left(\rho + p\right)\!u^a u_b + p\delta^a\!_b, \ee
with a linear equation of state $p=w \rho$ with $w$ constant. We begin by
reformulating the Replacement Principle for the stress-energy tensor of a
perfect fluid, as given in UW3 (see equations (84)),  by viewing it as a
function of the variables $F=(\rho,v_a,{\bar g}_{ab})$, where $v_a=a^{-1}u_a$
is the conformal fluid velocity and ${\bar g}_{ab}=a^{-2}g_{ab}$ is the
conformal metric tensor. The perturbations of the stress-energy tensor can be
written symbolically in the form:
\be \label{RP1}  {\cal M}^2\,{}^{(1)}\!{ T}^{a}\!_{b} =
\mathsf{T}^{a}\!_{b}(^{(1)}\!{F}),  \quad {\cal M}^2\,{}^{(2)}\!{T}^{a}\!_{b} =
\mathsf{T}^{a}\!_{b}(^{(2)}\!{F}) +
{\mathcal{T}}^{a}\!_{b}(^{(1)}\!{F}),    \ee
where $ \mathsf{T}^{a}\!_{b}$ is the linear leading order operator\footnote
{Note that the same operator $\mathsf{T}^a\!_b$ acts on both $^{(1)}\!{F}$ and $^{(2)}\!{F}$.}
 and
${\mathcal{T}}^{a}\!_{b}$ is the quadratic source term operator, and
\be {}^{(r)}\!F=({\cal M}^2\,{}^{(r)}\!\rho, {}^{(r)}\!v_a, {}^{(r)}\!{\bar g}_{ab}), \ee
with $r=1,2$, denotes the perturbations of $\rho, v_a$ and ${\bar g}_{ab}$,
while ${\cal M}$ is defined by
\be\label{cal_M}  {\cal M} := ({}^{(0)}\!\rho +{}^{(0)}\!p)^{-1/2}.  \ee
 For a linear equation of state the perturbations in
the pressure are given by
\be {}^{(r)}\!p=w\,{}^{(r)}\!\rho, \quad \text{for}\quad r=1,2; \ee

We associate gauge invariants with ${\cal M}^2\,{}^{(r)}\!{ T}^{a}\!_{b}$
and the variables ${}^{(r)}\!F$
as follows. The gauge invariants $^{(r)}\!A[X]$ associated with the perturbations
$^{(r)}\!A$ of an arbitrary tensor $A$ are defined by\footnote
{See Nakamura (2007), equations (2.26)-(2.27), and UW3, equations (81). Here
we are omitting the factor of $a^n$ in the latter equations.}
\begin{subequations}\label{A[X]}
\begin{align} {}^{(1)}\!{A}[X] &:= {}^{(1)}\!A
- \pounds_{{}^{(1)}\!X} {}^{(0)}\!A ,   \\
{}^{(2)}\!{A}[X] &:= {}^{(2)}\!A
- \pounds_{{}^{(2)}\!X}{}^{(0)}\!A
- \pounds_{{}^{(1)}\!X}\left(2A {}^{(1)} -
\pounds_{{}^{(1)}\!X}{}^{(0)}\!A\right).
\end{align}
\end{subequations}
In terms of this definition the gauge invariants associated with ${\cal
M}^2{}\,^{(r)}\!T^a\!_b$ and ${\cal M}^2{}\,^{(r)}\!\rho$ are defined by
replacing the tensor $A$ in~\eqref{A[X]} by $T^a\!_b$ and by $\rho$:
\begin{subequations}  \label{gi}
\be  \label{bdelta}   {}^{(r)}\!{\mathbb T}^a\!_b[X]:={\cal M}^2\, {}^{(r)}\!T^a\!_b[X],
\qquad {}^{(r)}\!{\bdelta}[X]:={\cal M}^2\, {}^{(r)}\!\rho[X],  \ee
while the gauge invariants associated with $^{(r)}\!v_a=a^{-1}{}^{(r)}\!u_a$
and ${}^{(r)}\!f_{ab}:={}^{(r)}\!{\bar g}_{ab}=a^{-2}\,{}^{(r)}\!{g}_{ab}$
are defined by replacing the tensor $A$ in~\eqref{A[X]} by $u_a$ and by
$g_{ab}$:
\be  {}^{(r)}\!{\bf v}_a[X]:=a^{-1}\, {}^{(r)}\! u_a[X], \qquad
{}^{(r)}\!{\bf f}_{ab}[X]:=a^{-2}\, {}^{(r)}\! g_{ab}[X].  \ee
\end{subequations}
The Replacement Principle  states that the gauge invariants associated with
${\cal M}^2\,{}^{(r)}\!T^a\!_b$ and with ${}^{(r)}\!F$ are related by {\it
the same} operators as in~\eqref{RP1}:
\begin{subequations}  \label{RP2}
\be  {}^{(1)}\!{\mathbb T}^{a}\!_{b}[X] =
\mathsf{T}^{a}\!_{b}(^{(1)}{\bf F}),\qquad {}^{(2)}\!{\mathbb T}^{a}\!_{b}[X] =
\mathsf{T}^{a}\!_{b}({}^{(2)}\!{\bf F}) +
{\mathcal{T}}^{a}\!_{b}({}^{(1)}\!{\bf F} ),    \ee
where ${}^{(r)}\!{\bf F}$ is shorthand for
\be  {}^{(r)}\!{\bf F}[X]= ({}^{(r)}\!\bdelta[X], {}^{(r)}\!{\bf v}_a[X], {}^{(r)}\!{\bf f}_{ab}[X]).  \ee
\end{subequations}

Finally the scalar velocity perturbations are defined by
\be \label{scalar_v}   ^{(r)}\!{\bf v}[X]={\cal S}^i \,{} ^{(r)}\!{\bf v}_i[X], \quad \text{for}\quad r=1,2, \ee
where the scalar mode extraction operator ${\cal S}^i$ was given
in~\eqref{extract}. We can now state that \emph{the first set of matter gauge
invariants for linear and second order scalar perturbations are
${}^{(r)}\!\bdelta[X]$ and $^{(r)}\!{\bf v}[X]$}, with $r=1,2$, relative to an arbitrary
choice of gauge.

Before continuing we briefly digress to comment on the relation between our
variables and ${}^{(r)}\!\bdelta[X]$ and $^{(r)}\!{\bf v}[X]$ and the
variables used by other authors. First, our ${\bf v}$ equals up to sign the
variable $v$ used by others.\footnote {To make comparisons we restrict to the
Poisson gauge ($X=X_{\mathrm p}$) at linear order. Noh and Hwang (2004) have
$v_{\chi}\equiv-{\bf v} [X_{\mathrm p}]$ (see equation (297)), Nakamura
(2007) has $v^{(1)}\equiv{\bf v} [X_{\mathrm p}]$ (see equation (5.20)) and
Malik and Wands (2009) have $v_{\ell}\equiv{\bf v} [X_{\mathrm p}] $ (see
equation (8.21)).} Second, as follows from~\eqref{cal_M} and~\eqref{bdelta},
our $\bdelta$ is related to the usual relative density perturbation $\delta
=^{(r)}\!\!\!\rho/^{(0)}\!\rho$ according to $\delta = (1+w){\bdelta}$, where
$w$ is defined by ${}^{(0)}\!p=w{}^{(0)}\!\rho$.

We also define matter gauge invariants directly in terms of the stress-energy
tensor:\footnote{The matter gauge invariants defined in UW3, using the
symbols $\Delta$ and $V$ instead of ${\mathbb D}$ and ${\mathbb V}$, are
related to the ones defined here as follows:
\be \label{old_new}  {}^{(r)}\!{\Delta}[X]={\cal A}^{(r)}\!{\mathbb D}[X],
\quad ^{(r)}\!{V}[X]={\cal A}^{(r)}\!{\mathbb V}[X]. \ee
For $\Delta$ see equations (45b) and (41c), and for $V$, see (45c).}
\begin{subequations} \label{bbVD}
\begin{align}
{}^{(r)}\!{\mathbb V}[X]&:={\mathcal S}^i\, {}^{(r)}{\mathbb T}^0\!_i[X],    \\
{}^{(r)}\!{\mathbb D}[X]&:=-{\mathcal S}^i \left({\bf D}_i{}^{(r)}{\mathbb T}^0\!_0[X] +
3{\cal H} ^{(r)}{\mathbb T}^0\!_i[X] \right).
\end{align}
\end{subequations}
In order to relate these gauge invariants to $^{(r)}\!{\bdelta}[X]$ and
$^{(r)}\!{\bf v}[X]$ we need to obtain explicit expressions for the
components of ${}^{(r)}{\mathbb T}^a\!_b[X]$. Referring to~\eqref{pf}
and~\eqref{RP1} we perform an $\epsilon$-expansion for $T^a\!_b(\epsilon),
g_{ab}(\epsilon), u_a(\epsilon)$ and $\rho(\epsilon)$ and obtain
\begin{subequations}   \label{Tab1}
\begin{alignat}{2}
{\mathsf T}^0\!_0(^{(r)}\!F)& = -{\cal M}^2\, {}^{(r)}\!\rho, &\quad  {\mathcal T}^0\!_0(^{(1)}\!F)&=-\gamma^{ij}v_{ij},  \\
{\mathsf T}^0\!_i(^{(r)}\!F) &= {}^{(r)}\!v_i, &\quad  {\mathcal T}^0\!_i(^{(1)}\!F)&=\left(2(1+w){\cal M}^2\,{}^{(1)}\!\rho +f_{00}\right)v_i, \\
{\mathsf T}^i\!_i (^{(r)}\!F)&= \,3w{\cal M}^2\,{}^{(r)}\!\rho, &\quad  {\mathcal T}^i\!_i(^{(1)}\!F)&=\gamma^{ij}v_{ij}, \\
{\hat{\mathsf T}}_{ij}(^{(r)}\!F)&=0, &\quad  {\hat{\mathcal T}}_{ij}(^{(1)}\!F)&=v_{ij},
\end{alignat}
where
 \be v_{ij}:=2\,v_i \left(v_j- f_{0j}\right),  \ee
\end{subequations}
and $r=1,2$ in the first column. We now apply the Replacement Principle
obtaining\footnote{For brevity, we omit the $[X]$ associated with the terms
on the right side of these equations.}
\begin{subequations}  \label{Tab2}
\begin{alignat}{2}
{\mathsf T}^0\!_0(^{(r)}\!{\bf F})& = -{}^{(r)}\!\bdelta, &\quad  {\mathcal T}^0\!_0(^{(1)}\!{\bf F})&=-\gamma^{ij}{\bf v}_{ij},  \\
{\mathsf T}^0\!_i(^{(r)}\!{\bf F}) &= -{}^{(r)}\!{\bf v}_i, &\quad
{\mathcal T}^0\!_i(^{(1)}\!{\bf F})&=\left(2(1+w)\bdelta +{\bf f}_{00}\right){\bf v}_i, \\
{\mathsf T}^i\!_i (^{(r)}\!{\bf F})&= \,3w\,{}^{(r)}\!\bdelta, &\quad  {\mathcal T}^i\!_i(^{(1)}\!{\bf F})&=\gamma^{ij}{\bf v}_{ij}, \\
{\hat{\mathsf T}}_{ij}(^{(r)}\!{\bf F})&=0, &\quad  {\hat{\mathcal T}}_{ij}(^{(1)}\!{\bf F})&={\bf v}_{ij},
\end{alignat}
where
\be {\bf v}_{ij}:=2\,{\bf v}_i \left({\bf v}_j- {\bf f}_{0j}\right).  \ee
\end{subequations}
When we specialize to linear perturbations that are purely scalar we
have\footnote{We recall that ${\bf f}_{00}[X]=-2\Phi[X]$ and ${\bf
f}_{0i}[X]={\bf D}_i {\bf B}[X] +{\bf B}_i$. See equations (8) in UW2.}
\be ^{(1)}\!{\bf v}_i[X]={\bf D}_i {}^{(1)}\!{\bf v}[X],  \qquad {\bf
v}_{ij}[X]:=2\,{\bf D}_i{\bf v}[X]\, {\bf D}_j\!\left({\bf v}[X]- {\bf
B}[X]\right) \ee
In this case it follows from~\eqref{RP2},~\eqref{bbVD} and~\eqref{Tab2} that the
two types of gauge invariants are related as follows:
\begin{subequations} \label{matter_relations}
\begin{align}
{}^{(1)}\!{\bf v}[X] &= {}^{(1)}\!{\mathbb V}[X], \\
{}^{(2)}\!{\bf v}[X] &= {}^{(2)}\!{\mathbb V}[X] - 2{\cal
S}^i\left[((1+w){\bdelta}[X] - \Phi[X]){\bf D}_i {\mathbb V}[X]  \right], \\
{}^{(1)}\!{\bdelta} [X] &= {}^{(1)}\!{\mathbb D} + {3\cH} ^{(1)}\!{\mathbb V}[X], \\
{}^{(2)}\!{\bdelta} [X] &= {}^{(2)}\!{\mathbb D}[X] + {3\cH} ^{(2)}\!{\mathbb
V}[X] - 2{\bf D}^i{\mathbb V}[X] {\bf D}_i({\mathbb V}[X] - {\bf B}[X]).
\end{align}
\end{subequations}
Finally in the Poisson gauge, which is given by ${\bf B}[X_{\mathrm p}]=0$,
equations~\eqref{matter_relations}, when specialized to dust simplify to give equations~\eqref{rel_gen}.

\end{appendix}

\section*{References}

\noindent	Bartolo, N., Matarrese, S., Pantano, O., and Riotto, A. (2010)	
Second-order matter perturbations in a $\Lambda$CDM cosmology and non-Gaussianity,
 {\it Class. Quant. Grav.} {\bf 27}, 124009. \\

 \noindent	Bartolo, N., Matarrese, S. and Riotto, A. (2006)
The full second-order radiation transfer function for large-scale CMB anisotropies,
{\it JCAP} {\bf 0605}, 010. \\

\noindent Bartolo, N., Matarrese, S. and Riotto, A. (2005)
Signatures of Primordial Non-Gaussianity in the Large-Scale Structure of the Universe,
{\it JCAP} {\bf 0510}, 010. \\

\noindent Boubekeur, L., Creminelli, P., Norena, J. and  Vernizzi, F. (2008)
Action approach to cosmological perturbations: the 2nd order metric in matter dominance,
{\it JCAP} {\bf 0808}, 028. \\


\noindent Bruni, M, Mena, F.C. and Tavakol, R. (2002) Cosmic no-hair:
non-linear asymptotic stability of de Sitter universe,
{\it Class. Quant. Grav.} {\bf 19}, L23-L29.\\

\noindent Goode, S.W. and Wainwright, J. (1985), Isotropic singularities
in cosmological models, {\it Class. Quant. Grav.} {\bf 2}, 99-115. \\

\noindent Hwang, J-C.,  Noh, H. and  Gong, J-O. (2012),
Second order solutions of cosmological perturbation in the matter dominated era,
{\it Astrophys. J.} {\bf 752}, 50, doi:10.1088/0004-637X/752/1/50.  \\

\noindent Malik, K. A. and Wands, D. (2009)
Cosmological perturbations, {\it Physics Reports} {\bf 475}, 1-51.\\

\noindent Matarrese, S., Mollerach, S. and Bruni, M. (1998)
Relativistic second-order perturbations of the Einstein-de Sitter
universe, {\it Phys. Rev. D} {\bf 58}, 043504 (1-22).\\

\noindent Mena, F.C., Tavakol, R. and Bruni, M. (2002) Second order
perturbations of flat dust FLRW universes with a cosmological constant,
{\it International Journal of Modern Physics A} {\bf 17}, 4239-4244. \\

\noindent 	Okouma, P. M., Fantaye, Y. and Bassett, B. A. (2013)
How flat is our Universe really? {\it Phys. Lett. B} {\bf 719} 1-4. \\
	
\noindent Noh, H. and Hwang, J. (2004) Second order
perturbations of the Friedmann world model, {\it Phys. Rev. D}
{\bf 69} ,104011(1-52). \\

\noindent Tomita, K. (1967) Non-linear theory of
gravitational instability in an expanding universe,
{\it Prog. Theor. Phys.} {\bf 37}, 831-846. \\

\noindent Tomita, K. (2005) Relativistic second-order
perturbations of nonzero-$\Lambda$ flat cosmological models and
CMB anisotropies,
{\it Phys. Rev. D} {\bf 71}, 083504(1-11). \\

\noindent Uggla, C. and Wainwright, J. (2011) Cosmological
perturbation theory revisited, {\it Class. Quant. Grav.} {\bf 28},
175017, arXiv:1102.5039.\\

\noindent Uggla, C. and Wainwright, J. (2012) Dynamics of
cosmological scalar perturbations, {\it Class. Quant. Grav.} {\bf 29},
105002, arXiv:1112.0880.\\

\noindent Uggla, C. and Wainwright, J. (2013) A simplified structure for the
second order cosmological perturbation equations, {\it Gen. Rel. Grav.} {\bf
45}, 643-674, DOI 10.1007/s10714-012-1492-7, arXiv:1203.4790.

\end{document}